\documentclass[twocolumn,showpacs,preprintnumbers,footinbib,prd,superscriptaddress,groupedaddress,10pt]{revtex4-1}
\usepackage[utf8]{inputenc}

\makeatletter
\def\l@subsubsection#1#2{}
\def\l@subsubsubsection#1#2{}
\makeatother

\usepackage{graphicx,amssymb,amsmath,amsthm,amsfonts}
\usepackage[usenames]{color}
\usepackage{mathtools}

\usepackage{notes2bib}
\usepackage{adjustbox}

\usepackage{aas_macros}
\usepackage{bm}
\usepackage{dcolumn}
\usepackage{latexsym}
\usepackage{rotating}
\usepackage{longtable}

\usepackage{enumerate}
\usepackage{tensor,multirow}
\usepackage{makecell,array}
\usepackage{url}
\usepackage[linktocpage]{hyperref}

\def\be{\begin{equation}}
\def\ee{\end{equation}}
\def\beq{\begin{eqnarray}}
\def\eeq{\end{eqnarray}}
\def\be{\begin{equation}}
\def\ee{\end{equation}}
	
\newcommand{\bea}{\begin{eqnarray}}
\newcommand{\eea}{\end{eqnarray}}

\begin{document}
\title{The light ring and the appearance of matter accreted by black holes}
\author{Vitor Cardoso}
\affiliation{CENTRA, Departamento de F\'{\i}sica, Instituto Superior T\'ecnico -- IST, Universidade de Lisboa -- UL,
Avenida Rovisco Pais 1, 1049 Lisboa, Portugal}
\author{Francisco Duque}
\affiliation{CENTRA, Departamento de F\'{\i}sica, Instituto Superior T\'ecnico -- IST, Universidade de Lisboa -- UL,
Avenida Rovisco Pais 1, 1049 Lisboa, Portugal}
\author{Arianna Foschi}
\affiliation{CENTRA, Departamento de F\'{\i}sica, Instituto Superior T\'ecnico -- IST, Universidade de Lisboa -- UL,
Avenida Rovisco Pais 1, 1049 Lisboa, Portugal}
\affiliation{Faculdade de Engenharia, Universidade do Porto, Rua Dr. Roberto Frias s/n, 4200-465 Porto, Portugal}
\begin{abstract} 
The geometry of black hole spacetimes can be probed with exquisite precision in the gravitational-wave window, and possibly also in the optical regime.
We study the accretion of bright spots -- objects which emit strongly in the optical or in gravitational waves -- by non-spinning black holes.
As the object approaches the event horizon, the emitted radiation to far-away distances is dominated by photons or gravitons orbiting the light ring, causing the total luminosity to
decrease exponentially as ${\cal L}_o\sim e^{-t/(3\sqrt{3}\,M)}$. Late-time radiation is {\it blueshifted}, due to its having been emitted during the infall, trapped at the light ring
and subsequently re-emitted.
These universal properties are a clear signature of the existence of light rings in the spacetime,
and not particularly sensitive to near horizon details.
\end{abstract}
%%%%
%%%%
\maketitle

%\tableofcontents

%%%%%%%%%%%%%%%%%%%%%%%
\section{Introduction}
%%%%%%%%%%%%%%%%%%%%%%%

%%%%%%%%%%%%%%%%%%%%%%%%%%%%%%%%%%%%%%
%\noindent{\bf{\em Introduction.}}
%%%%%%%%%%%%%%%%%%%%%%%%%%%%%%%%%%%%%%
The last few years have seen a substantial advance in our ability to probe compact objects, most notably black holes (BHs)~\cite{Barack:2018yly,Cardoso:2019rvt}.
The opportunities have grown considerably with the advent of gravitational-wave (GW) astronomy~\cite{PhysRevLett.116.061102,Abbott:2020niy}, and of precise electromagnetic observations with optical/infrared/radio very large baseline interferometry~\cite{Akiyama:2019cqa,Abuter:2020dou}. In the next few years, upgraded versions of Earth-based detectors, or new space-based missions~\cite{Akutsu:2018axf,Punturo:2010zz,Audley:2017drz} will improve our ability to study compact objects and test General Relativity with unprecedented precision over a wide range of scales.
Central to this new era, are the uniqueness results in General Relativity, which suggest that isolated BHs all belong to the same family of solutions -- the Kerr family~\cite{Kerr:1963ud} --
fully described by two parameters alone, mass and angular momentum~\cite{Chrusciel:2012jk,Robinson:2004zz}. To quote Chandrasekhar, BHs are ``the most perfect macroscopic objects there are in the universe''~\cite{Chandra}. This simplicity fixes the BH equation of state, and allows for a number of unique tests of gravity~\cite{Cardoso:2016ryw,Cardoso:2019rvt}.

For some of the physical processes involving BHs there is a notion of a stationary ``background'' spacetime, on which ``matter probes'' move and evolve.
These setups are particularly apt at providing detailed information of the geometry and underlying theory. In these circumstances, the geometry is effectively fixed and can be inferred by measuring accurately the motion of probes. 
Examples are ubiquitous. The gravitational multipole moments of the Earth, for example, can be determined in this way by studying the motion of orbiting satellites~\cite{tapley2004gravity,drinkwater2003goce,ciufolini2012overview}. In astrophysics, accretion flows around supermassive, otherwise isolated BHs, can also be well described as flows on a fixed Kerr background. The reason is simply that the matter content and density outside the BH is so small that its backreaction can be neglected for all practical purposes~\cite{Cardoso:2016ryw}. 
Thus, a fixed Kerr geometry is sufficient to understand and study the physics associated with observations by the Event Horizon Telescope~\cite{Akiyama:2019cqa} or GRAVITY~\cite{Abuter:2020dou}. The appearance of BHs, when illuminated by external sources, such as an accretion disk,
is of course dictated by those photons reaching far-away observers~\cite{1972ApJ...173L.137C,Luminet:1979nyg,Falcke:1999pj,Cardoso:2019dte,Gralla:2020srx,Cunha:2018acu}.
It is therefore no surprise that the separatrix between photons escaping to infinity and those eventually plunging into the BH horizon plays an important role in BH imaging. 
%In particular, photons which circle the BH an infinite number of times dictate the inner edge of a BH shadow. 
In particular, photons sent in from large distances with a decreasing impact parameter will be deflected with a larger angle, probing stronger-gravity regions before being scattered to observers far-away. Below a critical impact parameter, such photons simply fall onto the BH. At the critical value of impact parameter, the photon circles the BH an infinite number of times. 
These trajectories asymptote to a closed, unstable, photon orbit, which we will call the light ring. For non-rotating BHs, it is located at areal radius $3GM/c^2$.

The light ring thus controls the amount of information that one can gather, related to the BH geometry. But this fact concerns only matter external to the light ring itself. 
Here, we are interested in the appearance of luminous matter as it falls down a BH. Such events seem to occur periodically in the vicinities of the Sgr*A source~\cite{Baubock:2020dgq,Abuter:2020fpy}. 
Similar events were reported in the past in connection with the Cyg X-1 BH. In particular, dying pulses from BH accretion were discussed in the context of Cyg X-1, years ago~\cite{2001PASP..113..974D,2011arXiv1104.3164D}.
Emitters falling onto BHs may also be radiating in the GW window. These could be, for example, ``small binaries'' whose center-of-mass is inspiralling onto a supermassive BH~\cite{Cardoso:2021vjq}.
The dynamical appearance of bright sources was studied by Zeld'ovich and Novikov~\cite{Novikov:1965sik}, Podurets~\cite{1965SvA.....8..868P}, and Ames and Thorne~\cite{1968ApJ...151..659A}, but the analysis was based on a number of approximations and restricted to spherically symmetric gravitational collapse. Here, we investigate how a pointlike source, emitting GWs or electromagnetic waves, fades out as it is accreted by a BH.
The most salient feature is a universal exponential suppression of the total luminosity measured by far-away observers, ${\cal L}_o\sim e^{-t/(3\sqrt{3}\,M)}$, for all high-frequency radiators, due entirely to waves trapped at the light ring. The spectral content is dominated by radiation which was emitted with near-critical impact parameter and is {\it blueshifted}, $z\equiv \frac{\omega_e-\omega_o}{\omega_o} \sim -0.2$, with $\omega_e$ the proper frequency.
We also show that this universal ringdown is not very sensitive to near-horizon modifications, or to the details of the source inside the light ring.

Our numerical methods can be applied for a general Kerr geometry. However, for simplicity we focus exclusively on non-rotating BHs.
We use geometrical units where Newton's constant and the speed of light $G=c=1$.
%%%%%%%%%%%%%%%%%%%%%%%%%%%%%%%%%%%%%%%%%%%%%%%%%%%%%%%%%%%%%%%%%%%%%
\section{Light rings and photon surfaces: the key to compact objects}
%%%%%%%%%%%%%%%%%%%%%%%%%%%%%%%%%%%%%%%%%%%%%%%%%%%%%%%%%%%%%%%%%%%%%
Take a non-spinning BH of mass $M$ described in standard Schwarzschild $(t,r)$ coordinates (see Appendix~\ref{app_LR}).
Consider high-frequency radiation (light or GWs), described by null geodesics in this limit. In the Schwarzschild spacetime, the motion of null particles is encoded entirely in the radial equation~\cite{MTB,Cardoso:2008bp,Yang:2012he}
\beq
\dot{r}^2&=&E^2\left(1-f\frac{b^2}{r^2}\right)\equiv V\,,\\
f&=&\left(1-\frac{2M}{r}\right)\,,
\eeq
where $b$ is the impact parameter of the null particle, $E=f\dot{t}$ is a conserved energy parameter. Dots stand for derivative with respect to some affine parameter.
The above allows for a closed circular geodesic at $r=r_c$. Requiring that $\dot{r}=\ddot{r}=0\,\, (V=V'=0)$ one finds $b^2=r_c^2/f$ and $r_cf'-2f=0$ with prime a derivative with respect to radial coordinate. The solution is
\be
r_c=3M\,,\qquad b_{\rm crit}=3\sqrt{3}M\,.
\ee
This orbit has an angular frequency
%, measured by asymptotically far observers
$\Omega_{\rm LR}\equiv d\varphi / dt =1/(3\sqrt{3}M)$.
Such coordinate position defines, on the equatorial plane, a so-called light ring. Since this is a spherically symmetric spacetime, it defines more broadly a photon sphere.
This is the only closed null orbit outside the horizon.
Thus, high-frequency photons or GWs can be trapped at this location. It is an unstable trapping, since any small perturbation $\delta$ grows exponentially. For $r=r_c+\delta$,
\be
\delta \sim e^{\lambda t}\,,\label{timescale_LR}
\ee
with $\lambda=1/(3\sqrt{3}M)=\Omega_{\rm LR}$.
The relevant property is not only that high frequency radiation can be trapped at the light ring, but that it can do so from asymptotically large distances. In other words, high energy particles thrown from infinity with $b=b_{\rm crit}(1-\epsilon)$ will spiral round the light ring a number of orbits $n\propto -\log(\epsilon)$, on a timescale $\sim n/\lambda$ before plunging into the BH (see Appendix \ref{app_LR}).

Because of the above trapping properties and critical impact parameter for absorption onto a BH, light rings play a crucial role in our understanding of BHs.
The light ring is responsible for some of the relevant features of BH images~\cite{1972ApJ...173L.137C,Luminet:1979nyg,Falcke:1999pj,Cardoso:2019dte,Gralla:2020srx,Cunha:2018acu,Yang:2021zqy}.
Their trapping properties dictate the spacetime stability~\cite{Keir:2014oka,Cardoso:2014sna}.
It is the light ring ``relaxation'', described by Eq.~\eqref{timescale_LR} that controls the ``ringdown'' of BHs, the last stage of any fluctuation as probed by GW detectors~\cite{Cardoso:2008bp,Yang:2012he,Cardoso:2016rao,Cardoso:2016oxy,Cardoso:2019rvt}. They are, for many purposes, the inner surface probed by high-frequency observations. We now show, generalizing Podurets~\cite{1965SvA.....8..868P}, and Ames and Thorne~\cite{1968ApJ...151..659A}, that the light ring also dictates the late-time behavior of the luminosity of sources plunging into BHs.

%%%%%%%%%%%%%%%%%%%%%%%%%%%%%%%%%%%%%%%%%%
\section{How do bright bodies fade out?}
%%%%%%%%%%%%%%%%%%%%%%%%%%%%%%%%%%%%%%%%%%

%%%%%%%%%%%%%%%%%%%%%%%%%%%%%%%%%%%%%%%%%%
\subsection{An outward-pointing beam}
%%%%%%%%%%%%%%%%%%%%%%%%%%%%%%%%%%%%%%%%%%
%
\begin{figure}[htb]
\begin{tabular}{c}
\includegraphics[scale=0.46]{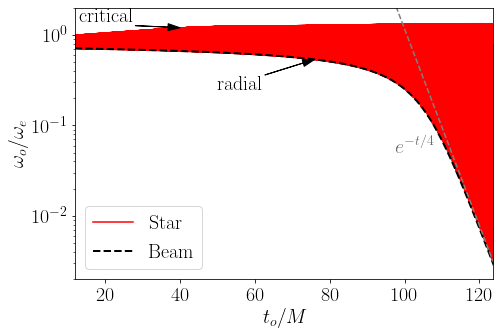}
\end{tabular}
\caption{
Redshift of two different sources as they plunge radially into a Schwarzschild BH, emitting photons or gravitons of fixed proper frequency.
The source, located in the equatorial plane at $\theta=\pi/2,\,\varphi=0$ begins from rest at infinity, but (for numerical purposes) starts emitting only when it crosses $r=30.65M$. 
{\bf Beam:} the first source (``beam,'' for example a laser pointer emitting a collimated beam of light) emits only radially. The observer is located at $r_o=100M$, $\theta=\pi/2$,
$\phi=0$, and receives photons whose energy decreases monotonically with time. At late times, the frequency measured by the $r_o=100M$ observer decays exponentially as $\omega_o\sim \omega_e e^{-t/(4M)}$. 
{\bf Isotropic star:} the second source is a pointlike ``star'' emitting isotropically in its local rest frame. At a fixed instant, far-away observers distributed along the celestial sphere at $r_o=100M$ receive a wide range of redshifts. The lower part of the curve is due to radially propagating null particles, whereas the top part of the curve is due to particles with a near critical impact parameter $b\approx 3\sqrt{3}M$ that linger close to the light ring. These can be blueshifted~\cite{Cardoso:2019dte}. The space in-between the two curves is ``naturally'' filled by the high number of photons that are collected (around 1 million photons), with varying redshifts. The early part of the curve is slightly dependent on initial conditions.
}
\label{fig:Infall_geometricoptics} 
\end{figure}
\begin{figure}[htb]
\begin{tabular}{c}
\includegraphics[scale=0.46]{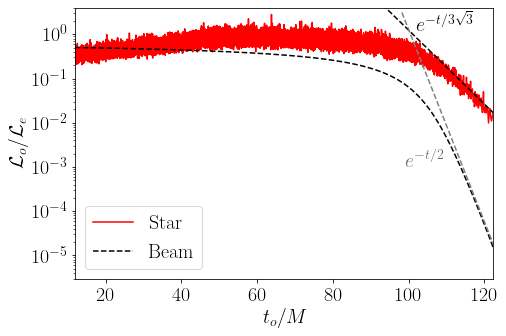}
\end{tabular}
\caption{
Normalized luminosity of the two different sources discussed in Fig.~\ref{fig:Infall_geometricoptics} above. There is no numerical noise here, only a scatter of photons of different energies reaching a sphere in the sky.
The observed luminosity of the radial beam scales as ${\cal L}\sim e^{-t/(2M)}$ at late times, in agreement with the prediction in the main text.
The luminosity of the isotropic star was calculated by (arbitrarily) ``binning'' null particles in packets of 20, in order to avoid large scatters. At late times, the luminosity is dominated by those particles lingering on the light ring, hence ${\cal L}\sim e^{-t/(3\sqrt{3} M)}$. Note that the late-time decay from these two sources is markedly different.
}
\label{fig:Infall_geometricoptics2} 
\end{figure}
To understand how a bright source fades out when it falls into a compact object, we focus first on the geometric-optics regime. In other words, we consider high-frequency sources
which can either be emitting in the GW or in the electromagnetic window. In this regime, the analysis can be done following null geodesics on a fixed background.

Consider first a collimated light source (say a laser pointer, which shoots a ``beam'' of light), freely falling, radially, from rest into a BH
and pointing radially outwards. In Schwarzschild coordinates, one finds the emitter four-velocity
\beq
v^\mu_e&=&(1/f,-x,0,0)\,,\label{radial_plunge}\\
x&\equiv&\sqrt{2M/r}\,,\quad f\equiv 1-2M/r
\eeq
In these conditions, the position of the laser pointer as function of proper time is 
\be
r_{e}=2M\left(-\frac{3\tau_e}{4M}\right)^{2/3}\,,\label{re_tau}
\ee
and the coordinate time at the laser pointer position is given implicitly by
\be
\frac{dt}{d\tau_e}=\frac{1}{f}\,.\label{te_tau}
\ee
Take a photon emitted at proper time $\tau_e$ with proper frequency $\omega_e=-(v_\mu k^\mu)_e$, with $k$ being the photon 4-momentum
and $v$ the laser pointer four-velocity. The photon is observed at event $o$ where it intersects the observer world-line, so that it is observed with
frequency $\omega_o=-(v_\mu k^\mu)_o$. For a static observer at large distances $v^\mu_o=(1,0,0,0)$. Now, for radial null geodesics one finds the momentum
\be
k^\mu=E(1/f,1,0,0)\,,\quad k_\mu=E(-1,1/f,0,0)\,.
\ee
We thus find
\be
\omega_o=\omega_e(1-x_e)\,.\label{redshift}
\ee

We now want to compute the redshift as seen by a distant observer, as function of time $t$. We need to take into account that the null particle 
is being emitted by a source which is getting closer to the horizon, and which also needs time to reach the observer.
An outward-directed photon obeys
\be
\frac{dt_{\rm travel}}{dr}=\frac{1}{f}\,.
\ee
We can integrate this to find the arrival time of the null particle as measured by a far-away observer
\be
t_o=t_e+(r_o-r_e)+2M\log\frac{r_o-2M}{r_e-2M}\,.\label{to_te}
\ee
Using Eqs.~\eqref{to_te} and~\eqref{redshift} we are now able to express both the ratio $\omega_o/\omega_e$ and the observer time $t_o$ as functions of the proper time $\tau_e$. Results are reported in Fig.~\ref{fig:Infall_geometricoptics}. At late times the redshift decreases as $\omega_o \sim e^{-t_o/4M}$. The same exponential behaviour can be obtained solving
\be
\frac{dr_e}{dt_e}=-\sqrt{\frac{2M}{r_e}}f_e\,,\label{dre_dte}
\ee
for sources close to the horizon. In fact, we find that  $t_e\sim -2M\log(r_e-2M)$ and using \eqref{to_te} one gets $r_e-2M\propto e^{-t_o/(4M)}$. For $r_e \simeq 2M$, $\omega_o \sim r_e -2M$ and hence we find a redshift 
\be
\omega_o\sim e^{-t_o/(4M)}\,,\label{redshift_laser}
\ee
at late times. The total luminosity $dE_o/dt_o$ can be calculated in a similar way.
At late times $dE_o/dt_o\sim e^{-t_o/(2M)}$.

Figures~\ref{fig:Infall_geometricoptics}-\ref{fig:Infall_geometricoptics2} show the numerical solution of this problem. An emitter starts falling at $r_i=30.65M$ and emits 20000 null particles, one every (proper) time interval 
$\delta \tau=4\times 10^{-3}M$, i.e. it corresponds to a monochromatic source of frequency $\omega_e = 2\pi/\delta \tau$. An observer at $r_o=100M$ collects these. 
Our numerical results show that at late times
the frequency as measured by far-away observer decreases exponentially as described by Eq.~\eqref{redshift_laser}. 
Note that the frequency measured by the observer at $r_o$ is always {\it redshifted}.
The luminosity, i.e., energy per second collected by this same observer, measured in units of proper luminosity, is shown
in the right panel. At late times, it falls exponentially as we just described.

%%%%%%%%%%%%%%%%%%%%%%%%%%%%%%%%%%%%%%%%%%%%%%%%%%%%%%%%%%%%%%%%%%%%%%
\subsection{An isotropically-emitting star}
%%%%%%%%%%%%%%%%%%%%%%%%%%%%%%%%%%%%%%%%%%%%%%%%%%%%%%%%%%%%%%%%%%%%%%
%
Most sources are not collimated. We can try to approximate infalling blobs of matter with an isotropically emitting,
pointlike source. Now, two other effects come into play: absorption at the horizon and, crucially, trapping close to the light ring.
Suppose that a luminous ``hot spot'' emits isotropically in its rest frame (where it has total luminosity ${\cal L}_e$)
while falling radially onto a BH. We now follow each emitted photon, or graviton to calculate total luminosity ${\cal L}_o$ as a function of time. Let us first compute the redshift.

\begin{figure}[t]
\begin{tabular}{c}
\includegraphics[scale=0.46]{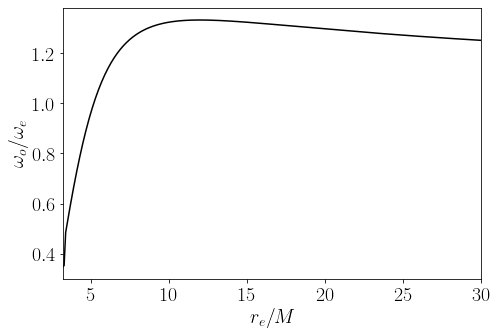}
\end{tabular}
\caption{The blueshift distribution of photons with near-critical impact parameter, emitted from a object freely-falling onto a BH. The blueshift is maximum, $\omega_o=4\omega_e/3$
at $r_e=12M$, and is unit at $r_e=3\sqrt{3}M$ (see also Ref.\cite{1982Ap&SS..88..307C}).
}
\label{fig:Redshift_absolute} 
\end{figure}
Null geodesics can be characterized by their energy $E$ and angular momentum $L$ at infinity via
\beq
p_\varphi&\equiv& r^2\dot{\varphi}=L\,,\\
-p_t&\equiv& f\dot{t}=E\,,\\
\dot{r}^2&\equiv&f (p_r)^2=E^2-f\frac{L^2}{r^2}=E^2A^2\,.
\eeq
Now, these null particles were emitted by a freely falling star, thus the conserved quantities $E, L$ must be related to observables in the locally freely falling frame,
where they have energy $\omega_e$ and are emitted at an angle $\alpha$ with respect to the radial direction (by symmetry, we focus on emission on the equatorial plane only).
Indeed, we can relate these quantities to locally-measured observables. One finds (see appendix for details)
\beq
\omega_o&=&\omega_e\left(1+x\cos\alpha\right)\,,\label{eq:redshift_star}\\
b\equiv \frac{L}{E}&=&r_e\frac{\sin\alpha}{1+x\cos\alpha}\,.\label{eq_b_alpha}
\eeq
We thus can now study the infall of an isotropic star by shooting null particles uniformly distributed in $\alpha$ and collecting them at some fixed radius $r_o$.
Those particles for which 
\be
b<b_{\rm crit}=3\sqrt{3}M\,,
\ee
will fall into the BH and are not considered in our calculation. 

Equations \eqref{eq:redshift_star}-\eqref{eq_b_alpha} can be solved for the redshift of particles with critical impact parameter. We find
\be
\frac{\omega_o}{\omega_e}=\frac{r_e^3+\sqrt{2M}\sqrt{r_e^5-b^2r_e^2(r_e-2M)}}{2Mb^2+r_e^3}\,,
\ee
The solution is shown in Fig.~\ref{fig:Redshift_absolute}. It shows that as the star falls, radiation with near-critical impact parameter is blueshifted for values of around $1.2-1.3$
for most of the fall (in particular, it is larger than 1.2 for $6.7M<r_e<49M$). The blueshift peaks at $r_e=12M$ and crosses unit at $r_e=b$,
\beq
\frac{\omega_o}{\omega_e}\biggr\rvert_{\rm max}&=&\frac{4}{3}\,,\quad r=12M\,,\\
\frac{\omega_o}{\omega_e}&=&1\,,\quad r=3\sqrt{3}M\,,
\eeq
in agreement with previous results~\cite{1982Ap&SS..88..307C}.

\begin{figure}[t]
\begin{tabular}{c}
\includegraphics[scale=0.46]{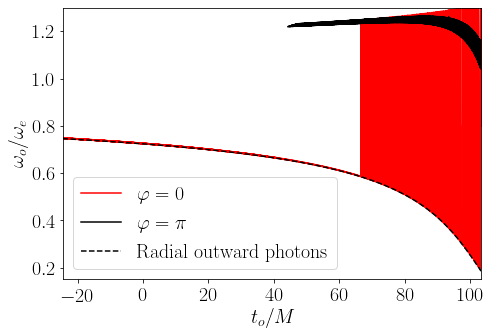}
\end{tabular}
\caption{The redshift distribution of light emitted by an infalling (isotropic) star as measured by observers at $r=100M$ on the infalling axis. For $\varphi=0$ the star is between the BH and the observer, and the observer initially sees mostly radially propagating particles only (the dashed line corresponds to the ``beam'' emission of Fig.~\ref{fig:Infall_geometricoptics}).
Observers at $\varphi=\pi$ only see the star due to gravitational lensing, as the BH sits between them and the star. Note the delay with which the $\varphi=\pi$ observer receives the first signal, with respect to $\varphi=0$. Note also the the signal is mostly Doppler blueshifted for $\varphi=\pi$, as the observer sees light emitted from an approaching source.
}
\label{fig:Star_Angle_Redshift} 
\end{figure}
We distributed 1600 particles uniformly in $\alpha$ and repeat the previous analysis. 
Our results are summarized in Figs.~\ref{fig:Infall_geometricoptics}-\ref{fig:Infall_geometricoptics2}.
The first important aspect is that light from an isotropically-emitting object (or other general source)
reaches far-away observers with a range of redshifts as Fig.~\ref{fig:Infall_geometricoptics} shows.
The lower part of the region agrees very well with the collimated ``beam'' curve. In other words, the most redshifted null
rays are those emitted radially.

The top part of the redshift region of shows that blueshifted rays can also be produced during infall.
As shown in Ref.~\cite{Cardoso:2019dte} these are a consequence of extreme bending, caused at or close to the light ring.
For example, a ray with a near-critical impact parameter can be bent by $\pi$, in other words it can be reflected back: we then have just a simple example of a moving source
and mirror. The ray is indeed expected to be blueshifted~\cite{Cardoso:2019dte}. Note that as the critical impact parameter is approached, the rays linger longer closer to the light ring, and take longer to reach the observer.

The redshift distribution discussed above refers to particles collected in the whole sky, at fixed radial coordinate $r_o=100M$. To understand what specific observers see, we selected
among all of the outgoing photons those that reach the observer with $\cos\varphi>0.99$ (which we label ``$\varphi=0$'') and those with $\cos\varphi<-0.99$ (which we label ``$\varphi=\pi$''). 
The corresponding distribution is shown in Fig.~\ref{fig:Star_Angle_Redshift} for observers at $r_o=100M$, and we remind that the star starts the infall at $r_e\sim 30M$. Observers with ``$\varphi=0$'' see the BH behind the star, and these three lie on the same axis.
Observers with ``$\varphi=\pi$'' see the star behind the BH.

As we can see the, at early times, ``$\varphi=0$'' observers see radially-moving null particles only. These are the particles that reach the observer first. However, after a sufficiently large amount of time, photons had time to circle the BH and also reach the observer. These photons are blueshifted. There is thus a ``phase transition'' here. Due to the finite opening angle, there's a varying redshift for the photons reaching this observer. Therefore, the thin red line in Fig.~\ref{fig:Star_Angle_Redshift} at early times represents the initial phase of this transition. In terms of actual observations, this initial stage will not be visible to an observer located at infinity, who will only see late-time phenomenology.
At late times,  particles with near-critical impact parameter circle the BH and come back to reach it. These should arrive a time $\Delta_1\sim T_{\rm LR}/2+60M\sim 76M$ after the first radially moving null particles, with $T_{\rm LR}$ being the time it takes to circle the light ring and come back in the opposite direction. On the other hand, an observer on the opposite side of the BH would see the first null particles to be always blueshifted, since the observer see a moving, approaching source, a time $\Delta_2 t\sim 60 M$ after the first signal arrives at the $\varphi=0$ observer. These estimates do not take into account Shapiro delay, but the estimate $\Delta_1-\Delta_2\sim T_{\rm LR}/2\sim 16M$ should be more reliable. All these features are apparent in Fig.~\ref{fig:Star_Angle_Redshift}.

The total luminosity ${\cal L}$ (flux integrated across sky) is shown in Fig.~\ref{fig:Infall_geometricoptics2} and follows the same trend. Note that due to the finite number of ``photons''
that we used in our numerical study, the total luminosity shown in Fig.~\ref{fig:Infall_geometricoptics2} is not smooth. The jagged features carry no physical information and are purely a
result of the numerical method used to estimate the luminosity. We opted to ``bin'' 20 particles at a time, and we have explicitly checked that larger binnings produce smoother luminosity functions, as it should. For realistic sources the true curve is single-valued and smooth, while our numerical approximation is thick and rough and approximates the real curve when the flux in the rest-frame increases.
At late times our results are consistent with a decay
${\cal L}\sim e^{-t_o/(3\sqrt{3}M)}$. Further details on this curve, in particular the luminosity per solid angle, are provided below for similar sources.

%%%%%%%%%%%%%%%%%%%%%%%%%%%%%%%%%%%%%%%%%%%%%%%%%%%%%%%%%%%%%%%%%%%
\subsection{An isotropic, scalar emitting body} \label{sec:Scalars}
%%%%%%%%%%%%%%%%%%%%%%%%%%%%%%%%%%%%%%%%%%%%%%%%%%%%%%%%%%%%%%%%%%%
%
\begin{figure}[htb]
\begin{tabular}{c}
\includegraphics[scale=0.46]{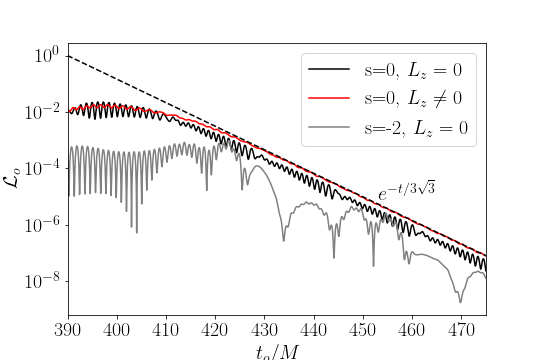}
\end{tabular}
\caption{
Total luminosity in scalar waves ($s=0$) and GWs ($s=-2$) from a source plunging into a Schwarzschild BH, and emitting at fixed proper frequency $M \omega_e = 2.5$. The source is located on the equatorial plane at $\theta = \pi/2,\, \varphi = 0$ and starts from rest at $r=35M$. We consider both a radial ($L_e=0$) plunge and one with finite angular momentum ($L_z\neq 0$, for which we fix $L_e=3.0M$). For both types of waves and plunging process, the luminosity follows the exponential decay dictated by the light ring ${\cal L}_o\sim e^{-t/(3\sqrt{3}M)}$ at late times, as seen in the geometric optics limit for the isotropic star in Fig.~\ref{fig:Infall_geometricoptics2}. The differences in features between scalar and GWs can be explained by the difference in the source structure. In particular, the low frequency oscillations in the GW spectrum are due to the plunge of the CM of the binary system (they are a consequence of the first 5 multipolar modes). The high-frequency content of the signal for both scalar and GWs is dominated by frequencies around $M\omega_o \sim 3.0$, higher than the proper frequency of emission by a factor $\sim 1.2$, consistent with these waves having been emitted during infall and trapped at the light ring, cf. Fig.~\ref{fig:Redshift_absolute}.
}
\label{fig:GWs} 
\end{figure}
We now consider a similar problem, dropping the geometric-optics approximation, and instead solving for the full dynamics of the field.
We start with a simple toy model that of a scalar charge $q$ (a scalar star) emitting scalar waves. For that, we solve the Klein-Gordon equation for a massless scalar field,
sourced by the trace of the stress-energy tensor of a pointlike body vibrating at constant proper frequency $\omega_e$ and thus emitting spherically symmetric waves in its rest frame~\cite{Nambu:2015aea,Nambu:2019sqn}
\be
\square \Psi = q\, T \, \sin \left(\omega_e \tau_e(t) \right)\,.\label{eq:KG} 
\ee
Denoting the worldline of the pointlike body by $z^\mu(\tau_e)=(t_e(\tau_e),r_e(\tau_e),\theta_e(\tau_e),\varphi_e(\tau_e))$, the explicit form of the stress-energy tensor is
% and respective trace is given by
%
\beq
T^{\mu\nu}(x)&=&m_e\int_{-\infty}^{+\infty}\delta^{(4)}(x-z(\tau))\frac{dz^\mu}{d\tau}\frac{dz^\nu}{d\tau}d\tau \label{eq:StressTensor}\\
&=&m_e\frac{dt}{d\tau_e}\frac{dz^\mu}{dt}\frac{dz^\nu}{dt}\frac{\delta(r-r_e(t))}{r^2}\delta^{(2)}(\Omega-\Omega_e(t))\,, \nonumber %\\
%T &=& g_{\mu\nu}T^{\mu\nu}(x) \, ,
\eeq
where $m_e$ is the mass of the body in its rest frame and we take $m_e\ll M$. 
All our results scale trivially with $m_e$ and $q$, and we therefore set these to unity henceforth.
In addition, we consider also sources with non-zero angular momentum $L_e$ along the equator. The motion is still described by conserved energy and angular momentum parameters for the source, whose radial motion can be studied solving numerically the differential equation
\beq
%\frac{d\tau_e}{dt} &=& f/E \, , \label{eq:tauCM}\\
\frac{dr_e}{dt} &=& -\sqrt{E_e^2 - f\left(1+ \frac{L_e^2}{r_e^2}\right)}  \, , \label{eq:rCM} %\\
%\frac{d\phi_e}{dt} &=& L/r_e^2\, , \label{eq:phiCM}\\
%\theta_e(t) &=& \frac{\pi}{2}\ \, . \label{eq:thetaCM}
\eeq

We solved Eq.~\eqref{eq:KG} using a time domain code that smoothens the pointlike source and was previously developed, tested and reported~\cite{Krivan_1997,LopezAleman:2003ik,Pazos_valos_2005,Sundararajan:2007jg, Cardoso:2021vjq}. 

The total luminosity for this system is shown in Fig.~\ref{fig:GWs} for a monochromatic object with $M\omega_e=2.5$, with and without angular momentum. Even though the source is now emitting radiation whose wavelength is comparable to the BH size, the late time behavior is still described by the exponential decay $\mathcal{L}_o \propto e^{-t_o/(3\sqrt{3}M)}$, independently of whether the body falls with non-zero angular momentum or not. 
\begin{figure}[t]
\begin{tabular}{c}
\includegraphics[scale=0.45]{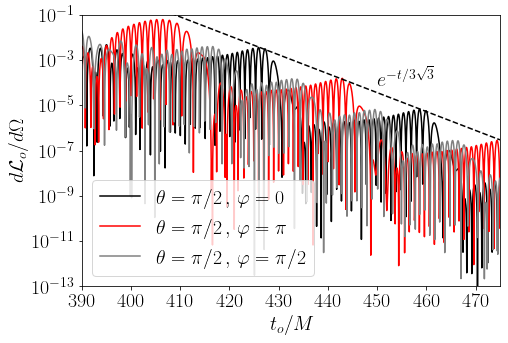}
\end{tabular}
\caption{Energy flux for a scalar source plunging radially into a BH (from $r_e=30M$ as in Fig.~\ref{fig:GWs}), extracted at specific angular positions on the equator. We are normalizing our results in units of $q m_e$. All signals exhibit the same global exponential decay dictated by the light ring as seen in Figs.~\ref{fig:Infall_geometricoptics2} and Fig.~\ref{fig:GWs}. Observers see now a periodic structure, whose period may differ for different observers (notice that at $\varphi=\pi/2$ the period is half that at $\varphi=0,\pi$). These features could mimic ``revivals'' reported in the literature~\cite{2001PASP..113..974D,2011arXiv1104.3164D}, but are independent of the motion of source, and rather related only to the light ring properties.  Once again, the high frequency content of the spectrum corresponds to waves with of $M\omega_o \sim 3.0$, in accordance with the blueshift predictions of Fig.~\ref{fig:Redshift_absolute}.
}
\label{fig:Comparison_Angular_Flux} 
\end{figure}

The luminosity per solid angle, at different angular positions, shows some more structure, as can be seen in Fig.~\ref{fig:Comparison_Angular_Flux}. The global ``light ring decay'' is the same. However, there are also periodic oscillations, whose period may differ for different observers. The frequency of these ``structures'' is a multiple of half of the frequency of the light ring $M\omega_{\text{LR}}\approx 0.192$ (corresponding to a period of $T_{\text{LR}}\approx 32.6M $). Each of these light ring pulsations is succeeded by a sharp, fast transition, lasting for $\sim5M$, a behavior and timescale that we do not fully understand.

An important component of this discussion concerns the spectral content of radiation at late times. As might be anticipated, the radiation is not low-frequency:
it is in fact dominated by blueshifted radiation that was emitted in the past with a near-critical angle. Referring to Fig.~\ref{fig:Redshift_absolute}, such radiation is blueshifted
to $\omega_o\sim 1.2-1.3 \omega_e$, in this case corresponding to $M\omega_o \sim 3.0-3.1$, during most of the infall. This is the radiation that the photon sphere absorbs to re-emit later. Our results agree with this line of argument. Late-time radiation is seen by asymptotic observers with frequency $M\omega_o \sim 3.0$.

%%%%%%%%%%%%%%%%%%%%%%%%%%%%%%%%%%%%%%%%%%%%%%%%%%%%%%%%%%%%%%%%
\subsection{A gravitational-wave emitting binary}\label{sec:GWs}
%%%%%%%%%%%%%%%%%%%%%%%%%%%%%%%%%%%%%%%%%%%%%%%%%%%%%%%%%%%%%%%%
%
\begin{figure}[t]
\begin{tabular}{cc}
\includegraphics[scale=0.46]{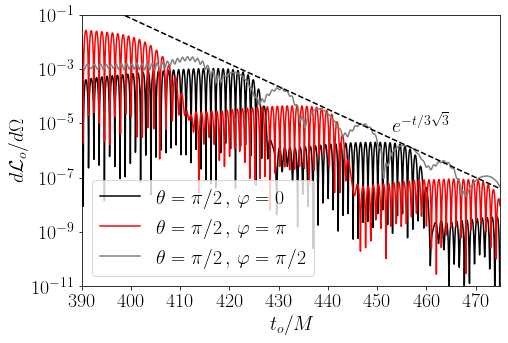}
\end{tabular}
\caption{Same as Fig.~\ref{fig:Comparison_Angular_Flux}, but for a source emitting high-frequency GWs. The source is a binary, and is plunging radially onto a massive BH, while emitting GWs of proper frequency $M \omega_e = 2.5$. The frequency of the signal measured by far away observers is blueshifted to $M\omega_o\sim 3$.}
\label{fig:Comparison_Angular_Flux_Grav} 
\end{figure}
Finally, we consider as source a binary system emitting GWs as it falls onto a massive BH. The binary is composed of two pointlike masses, small enough that the binary can be treated as a perturbation
on the background of the massive, central BH. Thus, they can also be represented by two pointlike particles with stress-energy tensor given by Eq.~\eqref{eq:StressTensor}, and the system constitutes a hierarchical triple~\cite{Cardoso:2021vjq}. Emission of GWs can be studied using Teukolsky's master equation ${\cal L}_s\Psi= \mathcal{T}$ \cite{Teukolsky:1973ha} where ${\cal L}$ is a second-order differential operator, $s$ refers to the ``spin weight'' of the perturbation field ($s=-2$ for GWs), and $\mathcal{T}$ is a spin-dependent source term~\cite{Teukolsky:1973ha} built from the stress-energy tensor \eqref{eq:StressTensor} (more details are given in Refs.~\cite{Sundararajan:2007jg,Cardoso:2021vjq}).

For the motion of the binary, we follow Ref.~\cite{Cardoso:2021vjq} and take the binary to be on a very-eccentric orbit around its center-of-mass (CM), while the CM itself is on a radial plunging trajectory, according to Eqs.~\eqref{radial_plunge}-\eqref{re_tau}. The motion of the binary around its CM can be parametrized by
\beq
r^\pm &=& r_\text{CM}(t) \, ,\qquad \theta^\pm = \theta_\text{CM}(t) \, ,  \\
\varphi^\pm &=& \varphi_\text{CM} + \epsilon \sin(\omega_e \tau_e)\, , 
\eeq
where $\pm$ refers to the two bodies composing the binary and $\epsilon=\epsilon(r_\text{CM})$ defines the axis of the very eccentric ellipse defined by the binary
\beq
\epsilon= (1-2M/r_\text{CM} ) \frac{\delta r }{r_\text{CM}} \, ,
\eeq
where $\delta r$ is the proper length of the axis of the binaries' motion around its CM. For the examples discussed here, we fix
$\delta r = 0.1M$.

Fluxes at different locations are shown in Figs.~\ref{fig:Comparison_Angular_Flux}-\ref{fig:Comparison_Angular_Flux_Grav}.
Notice that these are all systems emitting radiation, but the emission mechanisms and details vary. For example, it is impossible to decouple the CM-motion induced by GW emission from that
of the binary itself. Nevertheless, all these sources give rise to the same late time global exponential decay~${\cal L}_o\sim e^{-t/(3\sqrt{3}M)}$.
The peculiar nature of gravity is manifest on the low-frequency components in Fig.~\ref{fig:Comparison_Angular_Flux_Grav}: these are CM contributions. Superposed on these low-frequency components
we have the high frequency signal from the binary. 
Thus, for certain directions, such as $\theta=\pi/2$, $\phi = 0, \pi$, the high frequency contribution coming from the binary dominates the spectrum whereas for other angles the signal is controlled by the lower frequencies coming from the plunge of the CM.

%%%%%%%%%%%%%%%%%%%%%%%%%%%%%%%%%%%%%%%%%%%%%%%%%%%%%%%%%%%%%%%%%%%%%%
\section{Testing horizons}
%%%%%%%%%%%%%%%%%%%%%%%%%%%%%%%%%%%%%%%%%%%%%%%%%%%%%%%%%%%%%%%%%%%%%%
%
\begin{figure}[t]
\begin{tabular}{c}
\includegraphics[scale=0.45]{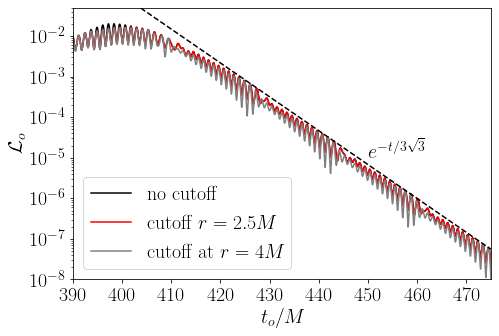}\\
\includegraphics[scale=0.46]{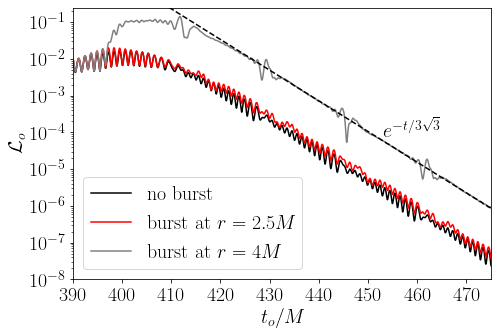}
\end{tabular}{}
\caption{\textbf{Top panel:} Luminosity in scalar waves for the system studied in Section~\ref{sec:Scalars}. Now, the source is turned off below a certain radius (which we selected to be either $r=2.5M$ or $r=4M$).
When the source is turned off inside or close to the light ring, the flux is nearly unchanged, as it is controlled by waves emitted in the past, and lingering close to the photon sphere.
\textbf{Bottom panel:} Luminosity for the scalar system studied in Section~\ref{sec:Scalars} but whose source is suddenly increased by a factor of 10 at the same radii as in top panel. In flat spacetimes, this would correspond to a luminosity 100 times higher.
However, since the process takes place close to the light ring, the luminosity is very weakly affected, and has the same global exponential decay. As expected, when the increase in amplitude occurs
deep inside the light ring, the increase in the luminosity is less significant.
}
\label{fig:Comparison_Scalars} 
\end{figure}
\begin{figure}[t]
\begin{tabular}{c}
\includegraphics[scale=0.46]{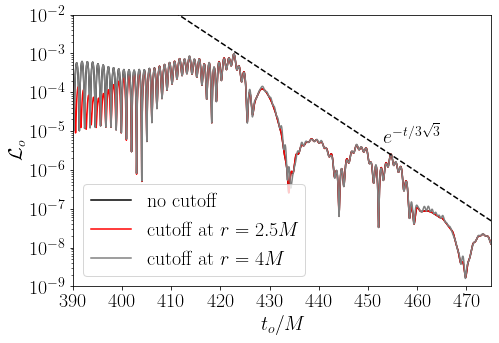}\\
\end{tabular}{}
\caption{Luminosity in GWs from the system described in Section~\ref{sec:GWs}, for which the binary is shut off below a certain radius (shown in the legend), signalling for example a sudden merger of the binary. In line with the findings for scalar waves in Fig.~\ref{fig:Comparison_Scalars}, near-horizon details are irrelevant for the appearance of these objects, and it is the light ring that controls the late time signal.
}
\label{fig:Comparison_GWs} 
\end{figure}
The previous results strengthen the point of view that the late-time appearance of BHs illuminated by matter is tightly connected to the photon sphere. Light rings control the way that dynamical processes
look like to an outside observer. One can take this program a step further and ask how these results change when the near horizon region changes, or equivalently if the source
is affected close to or within the light ring. We have studied two different types of processes:

\noindent {\bf (i)} in the first, we simply turn off the source term after it reaches selected radii, $r=4M,\, 2.5M$. This could signal, for example, a merging binary before its CM plunges on the BH. For scalar sources, it just stops shining. Results are summarized in Figs.~\ref{fig:Comparison_Scalars}-\ref{fig:Comparison_GWs}.
We observe that the spectrum, both for scalars and GWs, is only mildly dependent on the radius where the cutoff is performed, as long as it is close to or inside the photon sphere. Overall the flux of energy shows only a mild variation in amplitude. The late-time decay follows the exponential law dictated by the light ring seen in the previous sections. This fact strengthens the interpretation that the late-time decay really describes waves trapped close to the light ring, slowly leaking out. But this result also teaches us that two different compact objects with light rings may be hard to distinguish, even if one of them has an horizon and the other doesn't.

\noindent {\bf (ii)} the second change one can make is to let the source suddenly become brighter, increasing its proper luminosity after it falls within some radius.
The results of this experiment, when the proper luminosity is increased by a factor of 100, are shown in the lower panel of Fig.~\ref{fig:Comparison_Scalars}.
Again, the late-time decay is unchanged. Perhaps even more impressive is the only mild change in luminosity seen by far-away observers, specially for sources affected when they were already inside the photon sphere.

The results of this section indicate that it is the light ring and not near horizon details that are relevant for how matter accreted onto a BH appears to distant observers at late times.
``What happens inside the light ring stays inside the light ring'', is a tacky (and not formally correct) but effective lemma.
%%%%%%%%%%%%%%%%%%%%%%%%%%%%%%%%%%%%%%%%%%%%%%%%%%%%%%%%%%%%%%%%%%%%%%
\section{Discussion}
%%%%%%%%%%%%%%%%%%%%%%%%%%%%%%%%%%%%%%%%%%%%%%%%%%%%%%%%%%%%%%%%%%%%%%
The properties of BH light rings or photon spheres control, to a large extent, the appearance and late-time dynamics of BHs and other compact objects~\cite{1965SvA.....8..868P,1968ApJ...151..659A,1972ApJ...173L.137C,Luminet:1979nyg,Falcke:1999pj,Cardoso:2008bp,Yang:2012he,Cardoso:2016rao,Cardoso:2016oxy,Cunha:2018acu,Cardoso:2019rvt,Cardoso:2019dte,Gralla:2020srx,Yang:2021zqy}. The light ring can be seen as a trapping region, where high frequency waves are trapped on timescales~$\sim 3\sqrt{3}M$ or more~\cite{Cardoso:2019rvt}. A (GW or electromagnetic) bright source falling onto a BH will ``heat up'' this cavity as it falls. As the source enters the photonsphere, the transfer of energy from source to light ring is maximum. From then onwards, the source gets progressively redshifted away, and the energy leaking from the photon sphere dominates emission. Thus, observers see a late-time appearance of infalling stars dominated by the light ring cooling down process: the signal has a spectral content dominated by frequencies slightly blushifted with respect to the proper frequency of the source, and a luminosity dying off as ${\cal L}\sim e^{-t/(3\sqrt{3}\,M)}$.

This behavior is best understood within a null-particle approach. Literature on waves around BHs usually discusses, instead, a mode analysis where the late-time behavior is dominated by quasinormal ringdown and power-law tails~\cite{Leaver:1986gd,Berti:2009kk}. Gravitational waves, for example, are emitted by coherent motion of sources, and usually excite only a few of the modes. For high-frequency sources, however, a large number of multipoles are excited. The quasinormal frequencies at large mode number $\ell$, are described by~\cite{Berti:2009kk}
%
%\be
$\omega_{\rm QNM} =\Omega_{\rm LR}\left(\ell+1/2-i/2\right)\,.$
%\ee
%
In the ringdown stage the field amplitude is $\Phi\sim \sum_l e^{-i\omega_{\rm QNM} t}$. If we plug the asymptotic expression above in this sum over all the multipoles, we obtain a ringdown stage with a global modulation given by $\Phi \propto e ^{-\Omega_{\text{LR}}\, t/2 }$, which in the Schwarzschild case corresponds exactly to the decay in luminosity ($\mathcal{L} \propto |\Phi|^2$) we observed. In other words, both results (geometric optics and wave propagation) are compatible. At large frequencies it is perhaps more intuitive to use particle concepts (we should also note that, for similar reasons, a Doppler effect is also observed, and is the result of the sum of several multipoles~\cite{Cardoso:2021vjq}).
Finally, tails are extremely challenging to observe in the presence of these sources, as their amplitude is expected to be many orders of magnitude below the ringdown signal~\cite{Harms:2014dqa}. Consequently, they should only appear at later timescales than the ones probed in this work and for this reason are not expected to be astrophysically relevant.

\begin{figure}[t]
\begin{tabular}{c}
\includegraphics[scale=0.46]{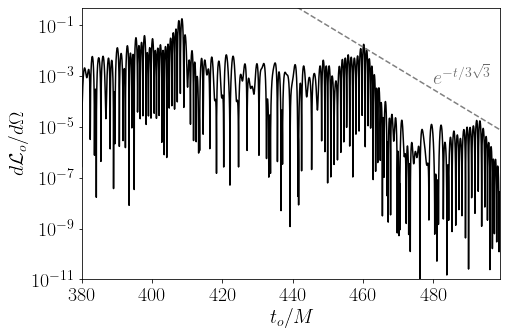}
\end{tabular}{}
\caption{Flux from a source emitting scalar waves on the innermost stable circular orbit of a BH, measured on the equator (and at $\varphi=0$). The source is suddenly turned off.
The late-time decay is the same as described in the main text, and corresponds to the ``cooling'' of the light ring, with the radiation that was fed onto it while the source was on circular motion.
The spectral content of the late-time signal is dominated by radiation blueshifted to $\omega_o\lesssim 1.4 \omega_e$ and redshifted down to $\omega_e\gtrsim 0.5 \omega_e$, as expected
from an analysis of the Doppler shift of radiation from such binaries~\cite{1972ApJ...173L.137C,2015MNRAS.448.2733C,Cardoso:2021vjq}. 
}
\label{fig:ISCO} 
\end{figure}
For a source falling from rest at infinity, the relaxation of the BH consists mostly of radiation which is blueshifted, with $\omega_o/\omega_e\sim 1.2$. 
The details will vary for sources with different energy, but the physics remains the same. Take for example a star orbiting close to the innermost stable circular orbit and which
then plunges into the BH. Such situation can be mimicked by letting a scalar source orbit on the innermost stable circular orbit (at $r=6M$) for a few orbits before being shut off.
This experiment is summarized in Fig.~\ref{fig:ISCO}.
While on circular motion, it feeds the light ring with blue- and red-shifted radiation (radiation which was emitted progradely and retrogradely). 
After the source is shut off (or plunging into the BH, as would happen to a star), the light ring re-emits this radiation. The relaxation timescale is the same as in the main text; the spectral content is now dominated by frequencies in a range dictated by redshifted and blueshifted radiation trapped by the light ring.

The decay timescale depends on the BH spin. We studied only non-spinning BHs, but geometric-optics approximation can be used to predict that rapidly spinning BHs will
show a much larger relaxation timescale, and a breaking of degeneracy with respect to different angular directions~\cite{Cardoso:2008bp,Yang:2012he}. This raises the interesting property
of determining the BH spin from the ratio of amplitudes of different redshifts, but requires significant further work.

Note that dying pulses from BH accretion were discussed in the context of Cyg X-1, years ago~\cite{2001PASP..113..974D,2011arXiv1104.3164D}.
These works assume that light from such pulses mimics the motion of the source, which as we discussed is not correct.
It is challenging to explain such observations through light ring properties, since timescales seem to be off by almost an order of magnitude;
nevertheless, they show how light ring relaxation could show up in observations with enough precision. 
Perhaps in the near future some of these aspects show up in observations of massive BHs. 

%%%%%%%%%%%%%%%%%%%%%%%%%%%%%%%%%%%%%%%%%%%%%%%%%%%%%%%
\noindent{\bf{\em Acknowledgments.}}
%%%%%%%%%%%%%%%%%%%%%%%%%%%%%%%%%%%%%%%%%%%%%%%%%%%%%%%
We thank the anonymous referees for the many useful suggestions that helped to improve the original manuscript.
V.C. acknowledges financial support provided under the European Union's H2020 ERC 
Consolidator Grant ``Matter and strong-field gravity: New frontiers in Einstein's 
theory'' grant agreement no. MaGRaTh--646597.
F.D. acknowledges financial support provided by FCT/Portugal through grant No. SFRH/BD/143657/2019. 
This project has received funding from the European Union's Horizon 2020 research and innovation programme under the Marie Sklodowska-Curie grant agreement No 101007855.
We thank FCT for financial support through Project~No.~UIDB/00099/2020.
We acknowledge financial support provided by FCT/Portugal through grants PTDC/MAT-APL/30043/2017 and PTDC/FIS-AST/7002/2020.
The authors would like to acknowledge networking support by the GWverse COST Action 
CA16104, ``Black holes, gravitational waves and fundamental physics.''
%
%%%%%%%%%%%%%%%%%%%%%%%%%%%%%%%%%%%%%%%%%%%%%%%%%%%%%%%%%%%%%%%%%%%%%%%

\appendix
%%%%%%%%%%%%%%%%%%%%%%%%%%%%%%%%%%%%%%%%%%%%%%%%%%%%%%%%
\section{Light ring relaxation properties\label{app_LR}}
%%%%%%%%%%%%%%%%%%%%%%%%%%%%%%%%%%%%%%%%%%%%%%%%%%%%%%%%
We focus on the Schwarzschild geometry written in standard coordinates,
\be
ds^2=-e^{2 \nu}dt^2+e^{-2 \nu}dr^2+r^2(d\theta^2+\sin^2\theta d\varphi^2)
\ee
with $e^{2 \nu}=f=1-2M/r$.

Take a null particle on the circular orbit at $r=r_c=3M$ and perturb it, so that $r=3M+\delta$.
The motion is controlled by $\dot{r}^2=V$. Expanding the potential close to the light ring, one finds
\be
\dot{\delta}^2=V(r_c)+V'(r-r_c)+\frac{(r-r_c)^2}{2}V''+...
\ee
By definition of circular orbit, the first two terms vanish. Thus, one gets
\be
\dot{\delta}^2=\frac{\delta^2}{2}V''\,.
\ee
Now $d\delta/d\tau=\dot{t}d\delta/dt$, therefore one can write
\be
\frac{d\delta/dt}{\delta}=\left(\frac{V''}{2\dot{t}^2}\right)^{1/2}\,,
\ee
with solution 
\beq
\delta&\sim& \delta_0 e^{\lambda \,t}\,,\\
\lambda&=&\left(\frac{V''}{2\dot{t}^2}\right)^{1/2}=\left(\frac{V'' f^2}{2E^2}\right)^{1/2}=\frac{1}{3\sqrt{3}M}\,.
\eeq

In other words, a null ray slightly displaced off the light ring will orbit on a timescale $t\sim \log \delta /\lambda$.
During this timescale, the null particle does a number of orbits
\be
n\sim \frac{\Omega t}{2\pi}=-\frac{\log{\delta}}{2\pi} \,,
\ee
close to the LR. For further details and refined estimates see Chandrasekhar's classical work~\cite{MTB}.

%%%%%%%%%%%%%%%%%%%%%%%%%%%%%%%%%%%%%%%%%%%%%%%%%%%%%%%
\section{An isotropically-emitting star\label{app_iso}}
%%%%%%%%%%%%%%%%%%%%%%%%%%%%%%%%%%%%%%%%%%%%%%%%%%%%%%%
Here, we provide some details on the calculation of the emission of isotropic stars. For that we need to describe the physics as seen by a freely-falling observer.
The following builds on - and agrees with - Refs.~\cite{Misner:1974qy,1975PhRvD..12..323C,Siwek:2015dqa}.

%Consider the appearance of the external universe as seen by two different observers: a static observer, i.e. observer at rest in the external field of the BH ($r=\theta=\varphi= \rm const.$). 
%We focus on freely-falling observers who start from rest at spatial infinity (radial geodesic motion). The extension to other cases is trivial.  

Let consider two different observers: a static observer, i.e.  characterized by a wordline with $r=\theta=\varphi= \rm const.$ and a free-falling observer, who starts from rest at spatial infinity and has a purely radial motion.

Start first with an observer at rest on the equatorial axis ($\{r, \theta, \varphi \} = \{ r_e, \pi/2, \rm arbitrary \}$) in a proper reference frame with basis
\be
\omega^{\hat{t}} = e^{\nu} dt\,,\quad \omega^{\hat{r}} = e^{-\nu} dr\,,\quad \omega^{\hat{\varphi}}=r d\varphi\,.\label{static_basis}
\ee

If we consider a photon emitted by a source at rest at infinity and received by the observer, its geodesic motion is fully determined by its energy $E$ and its impact parameter $b$.  The components of the photon's four momentum read:
%Consider a photon emitted by a source at rest at infinity and received by the observer. The photon's worldline is characterized by its energy $E$ and its impact parameter $b$.  
%
\beq
p_t&=&-e^{2 \nu} \dot{t}= -E \,,\quad p_{\varphi} = r^2 \dot{\varphi} = L \equiv  b E\,,\nonumber \\
%
%p_{\varphi}&=&r^2 \sin^2 \theta \dot{\varphi} = 0\,,\nonumber \\
%
p_r&=& e^{-2 \nu} \dot{r} = A \, e^{-2 \nu} E \,,\quad A^2 \equiv 1-b^2 r^{-2} e^{2 \nu}\,,\label{four_momentum}
\eeq
%
%Since we are considering $\theta = 0$, the conserved angular momentum is associated with the $\theta$-component of the metric, in fact 
%
%\be
%l=\vec{r} \wedge m \vec{v} = m r^2 (\dot{\theta} \mathbf{\hat{\varphi}} - \dot{\varphi} \sin \theta \mathbf{\hat{\theta}})
%\ee

We must now compute the $p^{\hat{t}}$ component of the momentum in the observer's reference frame. From (\ref{static_basis}) we get 
\be
dt=e^{-\nu} \omega^{\hat{t}}\,,\quad dr=e^{\nu} \omega^{\hat{r}}\,,\quad d\varphi = r^{-1} \omega^{\hat{\varphi}}\,,
\ee
and hence 
%
%\be
%\begin{pmatrix} 
%-p^{\hat{t}} \\ p^{\hat{r}} \end{pmatrix} = \begin{pmatrix} e^{-\nu} & 0 \\
%0 & e^{\nu} \end{pmatrix} \begin{pmatrix}  p_t \\ p_r 
%\end{pmatrix}
%\ee
%
%
\be
p^{\hat{t}} = - e^{-\nu} p_t \,,\quad p^{\hat{r}} = e^{\nu} p_r = A\, E \, e^{-\nu}\,.
\ee

The ratio of observed to emitted energy is then
\be
\frac{p^{\hat{t}}}{E} = e^{-\nu}\,. \label{blueshift}
\ee
%
%if we consider a bunch of photons with almost the same energy and impact parameter, the ratio of observed to emitted specific intensities is 
%\begin{equation}
%\frac{I_{\nu_0}}{I_{\nu_e}}=\left( \frac{p^{\hat{t}}}{E} \right)^3 = e^{-3 \nu}
%\label{ratio_intensities}
%\end{equation}
%
%This relation comes from the fact that the number density $\mathcal{N} \equiv N/V$ in phase space is a conserved quantity (Liouville's theorem in curved spacetime). For photons is possible to obtain %an analogous result considering the specific intensity $I_{\nu}$ instead of $\mathcal{N}$. We have
%\begin{equation}
%\mathcal{N} = h^{-4} \left(\frac{I_{\nu}}{\nu^3} \right)
%\label{intensity}
%\end{equation}
%
%Thus, if two different observers at the same or different events in spacetime look at the same photon as it passes them, they will see different frequencies $\nu$, different specific intensities %$I_{\nu}$ but they will obtain identical values for the ratio $I_{\nu}/\nu^3$. Hence
%\begin{equation}
%\frac{I_{\nu_0}}{h E_{\nu_0}^3} = \frac{I_{\nu_e}}{h E_{\nu_e}^3} \,\,\,\, \rightarrow \,\,\, \frac{I_{\nu_0}}{I_{\nu_e}} = \left( \frac{E_{\nu_0}}{E_{\nu_e}} \right)^3  
%\end{equation}

Moreover, the observer sees the null rays come in at an angle $\alpha$ relative to its radial direction given by
\be
\cos \alpha = -\frac{p^{\hat{r}}}{p^{\hat{t}}}= -A\,.
%& = \pm \left[1 - \left(\frac{b}{r_0}\right)^2 \left(1-\frac{2m}{r_0} \right) \right]^{1/2}
\label{cos_alfa}
\ee
Here, $\alpha$ is the angle between the propagation direction and the radial direction and it is defined as $\cos \alpha = v_{\hat{r}}$, where $v_{\hat{r}}$ is the velocity of the massless particle relative to the observer's reference frame,
\begin{equation}
v_{\hat{r}} = \frac{|g_{rr}|^{1/2} dr/d\lambda}{|g_{00}|^{1/2} dt/d \lambda}\,.
\end{equation}

Consider now free-falling observers. The basis one-forms of their proper reference frame are
\be
\omega^{\hat{t}} = dt + x e^{-2 \nu} dr\,,\quad \omega^{\hat{r}} = x dt + e^{-2 \nu} dr \,, \label{basis_freefall}
\ee
with $x=(2m/r_e)^{1/2}$.  When a photon with energy at infinity $E_o$ and impact parameter $b$ reaches the observer at $r=r_e$ and $\theta = \pi/2$, its four momentum is given by \eqref{four_momentum}. On the other hand,  infalling observers will see the photon comes in with an energy  $p^{\hat{t}} = \omega^{\hat{t}} \cdot \mathbf{p}$ and an angle $\alpha = \cos^{-1} (-p^{\hat{r}}/p^{\hat{t}})$ to the radial direction. 
%Take a photon with energy at infinity $E_o$ and impact parameter $b$. When it reaches an infalling observer at $r=r_e$ and $\theta = \pi/2$, the photon has four momentum given by \eqref{four_momentum}.  The observer sees the photon to have energy $p^{\hat{t}} = \omega^{\hat{t}} \cdot \mathbf{p}$ and to come in with an angle $\alpha = \cos^{-1} (-p^{\hat{r}}/p^{\hat{t}})$ to the radial direction. 
As before, using (\ref{basis_freefall}) we get 
\be
dt = \frac{-\omega^{\hat{t}} + \omega^{\hat{r}} x}{x^2 -1} \,,\quad  dr = \frac{e^{2\nu}(-\omega^{\hat{r}}+\omega^{\hat{t}} x)}{x^2 -1}\,.
\ee
%
%In order to obtain the 4-momentum in the new basis we must compute: 
%\begin{equation}
%\begin{pmatrix} 
%-p^{\hat{t}} \\ p^{\hat{r}} \end{pmatrix} = \begin{pmatrix} - \frac{1}{x^2 -1} & e^{2 \nu} \frac{x}{x^2 -1} \\
%\frac{x}{x^2 -1} & - \frac{e^{2 \nu}}{x^2 -1} \end{pmatrix} \begin{pmatrix}  p_t \\ p_r \end{pmatrix}
%\end{equation}
Hence, we recover the results of Ref.~\cite{1975PhRvD..12..323C}
%
%\af{Don't we want to add also
\beq
\cos\alpha &=& - \left(\frac{p^{\hat{r}}}{p^{\hat{t}}} \right) =  - \frac{x+a}{1+a x}\,, \\
\frac{p^{\hat{t}}}{E} &=& - \frac{1+ax}{x^2-1} = \frac{1}{1+ x \cos \alpha} \,,\label{shift_appendix}\\
%\eeq, that are the results we use in the main text? In this case, I'd keep only (B14) in the following equations}
%
%\beq
%\cos^2 \alpha &=& 1 - \left(\frac{b}{r_e} \right)^2 e^{2 \nu}\,, \\
%\sin^2 \alpha &=& \left(\frac{b}{r_e}\right)^2 e^{2 \nu}\,, \\
\frac{b}{r_e} &=& \sin \alpha \, e^{-\nu} = \sin \alpha \left(\frac{p^{\hat{t}}}{E} \right) = \frac{\sin \alpha}{1 + x \cos \alpha}\,.
\eeq
%
%\af{Note that in the main text we use the inverse relation with respect to  \eqref{shift_appendix} because we are interested in the frequency-shift of a free-falling emitter seen by a distant observer}
%%
%\beq
%-p^{\hat{t}}&=& -\frac{p_t}{x^2-1} + e^{2 \nu}\frac{x}{x^2 -1} p_r = \frac{E(1+a x)}{x^2-1} \\
%%
%p^{\hat{r}}&=& \frac{x}{x^2 -1} p_t - e^{2 \nu} \frac{p_r}{x^2 -1} = - \frac{E(a+x)}{x^2 -1}\,,
%\eeq
%%
%and
%%
%\be
%\cos\alpha = - \left(\frac{p^{\hat{r}}}{p^{\hat{t}}} \right) =  - \frac{x+a}{1+a x}\,.\label{photons_obs}
%\ee
%
%\begin{equation}
%\frac{p^{\hat{t}}}{E} = - \frac{1+ax}{x^2-1} = \frac{1}{1+ x \cos \alpha}
%\end{equation}
%
%\noindent We can get Eq.(14.c) in Cunningham's paper  from Eq.(\ref{cos_alfa}):  
%

\bibliographystyle{h-physrev4}
\bibliography{references}

\end{document}